\documentclass[a4paper,11pt]{article}
\usepackage{jcappub} 
\usepackage{lineno}
\usepackage{wasysym}

\title{\boldmath Revisiting astrophysical bounds on continuous spontaneous localization models}

\author[1,2]{M. M. Ocampo}
\author[1,2]{M. M. Miller Bertolami}
\author[2,3]{G. León}
\affiliation[1]{Instituto de Astrofísica de La Plata, CONICET - Universidad Nacional de La Plata\\
 Av. Centenario S/N, Paseo del Bosque, La Plata, Argentina}
\affiliation[2]{Facultad de Ciencias Astronómicas y Geofísicas, Universidad Nacional de La Plata\\
 Av. Centenario S/N, Paseo del Bosque, La Plata, Argentina}

\affiliation[3]{CONICET, Godoy Cruz 2290, 1425 Ciudad Aut\'onoma de Buenos Aires, Argentina}

\emailAdd{mocampo@fcaglp.unlp.edu.ar}

\abstract{Among the open problems in fundamental physics, few are as conceptually significant as the measurement problem in Quantum Mechanics. One of the proposed solutions to this problem is the  Continuous Spontaneous Localization (CSL) model, which introduces a non-linear and stochastic modification of the Schrödinger equation. This model incorporates two parameters that can be subjected to experimental constraints. One of the most notable consequences of this theory is the spontaneous heating of massive objects; this anomalous heating is dependent on the CSL parameters. In this work, we will revisit some astrophysical bounds previously found, and introduce new methods for testing the spontaneous heating in a variety of compact objects. Finally, we will compare our different bounds and discuss the benefits and shortcomings of each one.}

\begin{document}
\maketitle
\flushbottom

\section{Introduction}
\label{sec:intro}



The measurement problem in Quantum Mechanics (QM) \cite{Wigner63, Omnes, Bell87, Bell1995, Albert, Okon2014, Norsen2017, Becker} arises from the coexistence of two different rules for the time evolution of a system: the deterministic dynamics dictated by Schrödinger's equation, and the so-called \textit{collapse} of the wave function. This collapse represents an abrupt and random change in the system's physical state during measurement, contrasting with the continuous and deterministic evolution described by Schrödinger's equation. Specifically, the act of measurement appears to break or ``collapse'' the superposition of states (as allowed by Schrödinger's equation) into a single definite state;\footnote{The change in the state of the physical system caused by the act of measurement also raises profound questions about the role of the observer and the nature of reality. What physical systems qualify to play the observer's role? Can QM also be used to characterize an observer or measurement device?} the latter is what is always observed in actual measurements. However, the theory does not unambiguously and precisely define what constitutes a measurement. In other words, QM does not specify exactly when to apply the collapse postulate instead of Schrödinger evolution.

The measurement problem has been known since the dawn of QM and is often treated as of purely philosophical nature, essentially because of the enormous success the theory has enjoyed in applications, ranging from particle physics to condensed matter. Nevertheless, as J.S. Bell has argued \cite{Bell87, Bell1995}, such a pragmatic attitude is sometimes inadequate; for instance, when applying QM to the early universe \cite{Hartle1993,perezsahlmansudarsky2006,Sudarsky:2009za,Landau:2011ljv,Sudarsky2020,Bengochea2020b,Leon2021,Piccirilli2023}. 
Maudlin has characterized the quantum measurement problem in a formal manner \cite{Maudlin1995-MAUTMP}. There,  it is shown that the following three premises cannot be held simultaneously in a self-consistent manner: 
\begin{enumerate}
    \item The physical description provided by the state vector is complete. \label{A}
    \item Quantum states always evolve according to the Schrödinger equation. \label{B}
    \item Results of experiments, i.e. measurements, always have definite results. \label{C}
\end{enumerate}

Objective collapse theories \cite{GRW86,GRW87,Pearle1976, Diosi:1984wuz,Diosi:1986nu,Diosi:1988uy,penrose1989book},  also known simply as \textit{collapse models}, are a proposal to address the measurement problem by neglecting premise \ref{B}. Collapse models postulate that each physical system interacts with an universal noise field, triggering the collapse of the wave function in space. In this scenario, the evolution of the system is given by a modified  Schrödinger equation that incorporates the effect of the noise field. The effect of the noise is almost negligible for microscopic systems but becomes dominant for macro systems; this is known as the amplification mechanism. In this manner, the aim of these models is to extend QM in order to obtain a theory that reproduces its successful predictions in the microscopic realm, and can also be applied to describe macroscopic phenomena. In this work, we will analyze the Continuous Spontaneous Localization (CSL) model \cite{Pearle1976,Pearle1989, Ghirardi1989, Ghirardi1989b}. The CSL model is characterized by two parameters: a collapse rate, $\lambda$ (typically very small $\sim 10^{-16}$ s$^{-1}$, ensuring rare collapses for microscopic systems), and a localization length, $r_C$ (typically small $\sim 10^{-7}$ m, ensuring rapid localization for macroscopic systems).
The CSL model has also been subject to experimental testing as its predictions differ from those of standard
QM for certain systems \cite{Gasbarri2021,BassiNat}. These empirical constraints have been obtained from various experiments, including: spontaneous X-ray emission \cite{sandro2017,Majorana2022}, matter–wave interferometry \cite{toros2016} and gravitational
wave detectors \cite{carlesso2016, PhysRevD.95.084054}.  

Since Newton proposed that the laws of motion that governed the stars and objects terrestrials were the same, the stars have been used multiple times to test new theories in physics. The last decades have not been an exception from this tradition. The  need to incorporate a cosmological constant in our description of the universe was initially provided by the study of the luminosities of SNeIa \cite{1998AJ....116.1009R, 1999ApJ...517..565P}, and the problem of solar neutrinos offered a clear example of how stars can be used for detecting the need for new physics \cite{1968PhRvL..20.1205D,1998ApJ...496..505C,2005ApJ...621L..85B, 2006ApJS..165..400B}. Moreover, astrophysical arguments are routinely used to put constraints on either new hypothetical particles and fields or on hypothetical properties of well-established particles  \cite{1996slfp.book.....R, 2016JCAP...05..057G, 2017PhRvD..96k5021K}. Likewise, the deviations from standard physics predicted by the CSL model can also be constrained by astrophysical arguments. 

A novel physical prediction of collapse models is spontaneous heating due to the interaction of the noise field with the system (i.e. there is a measurable non-conservation of energy). This mechanism provides a tool for constraining the parameters of the CSL model  \cite{Vinante2016, Carlesso2016bb, Helou2016, Vinante2016b, Carlesso2017, Mishra2018, vinante2018, mohammad2018}. In Refs. \cite{Tilloy2019, Adler2019}, the CSL model was applied to a Fermi liquid, obtaining an expression for the heating power induced by the collapse mechanism. Using this expression the authors made a rough estimation for the value of $\lambda/{r_C}^2$ above which observable consequences in the structure and evolution of neutron stars and planets should appear. This was done by assuming that the thermal radiation coming from the surface of those astrophysical objects was balanced by  the anomalous heating predicted in the CSL theory.
In this work we will extend this analysis to a variety of astrophysical objects. The paper is organized as follows: In Section \ref{sec2}, we will outline the basics of the Continuous Spontaneous Localization (CSL) model and its connection to the resulting heating effect.
In Section \ref{sec3}, we will derive bounds from the physics of white dwarfs, using both white dwarf populations and individual stars. Then, in Section \ref{sec_planet_revisited}, we will revisit the bounds derived from the internal heat flow of planets in the Solar System, and show that when external heating from the Sun is taken into account, these bounds are as restrictive as most laboratory experiments. In Section \ref{secneutron}, we will revise the bounds from neutron stars by focusing on an interesting neutron star candidate. Finally, in Section \ref{discussion}, we present our results and discuss the benefits, shortcomings, and prospects of the different astrophysical bounds on the CSL parameters.

\section{CSL heating} \label{sec2}

In this section, we will summarize the fundamentals of the Continuous Spontaneous Localization (CSL) model and its relation to the induced heating effect. Therefore, no original work is presented in this section. A complete review of the theory (including its derivation) can be found in \cite{2003PhR...379..257B, Bassi2012bg, Pearle2012}.  

The CSL model is an extension of Quantum Mechanics, in which a non-linear stochastic term is added to the Schrödinger equation. The non-linearity serves to break quantum superpositions, providing a self-induced collapse of the wave function in position space. Stochasticity, on the other hand, prevents faster-than-light signaling. The model also recovers the probability rule given in Born's postulate. 

The dynamics of the (mass proportional) CSL model is given by the following stochastic differential equation
\begin{eqnarray}\label{CSLmaster}
    d |\psi (t) \rangle &=& \bigg\{  - \frac{i}{\hbar} \hat H dt       + \frac{\sqrt{\gamma}}{m_0} \int d^3 \textbf{x} [\hat M(\textbf{x}) - \langle \hat M(\textbf{x} ) \rangle ] dW(\textbf{x},t) \nonumber \\ 
    &-& \frac{\gamma}{2 m_0} \int d^3 \textbf{x} [\hat M(\textbf{x}) - \langle \hat M(\textbf{x} ) \rangle ]^2 dt \bigg\} |\psi (t) \rangle,
\end{eqnarray}
where $\gamma \equiv \lambda  r_C^3  8 \pi^{3/2} $, $\hat H$ is the standard Hamiltonian of the system, $\langle \hat M \rangle \equiv \langle \psi (t) | \hat M | \psi (t) \rangle$, $m_0$ is a reference mass, traditionally chosen as the mass of a nucleon,  and $W (\textbf{x},t)$ is an ensemble of independent Wiener processes, one for each point in space. The operator $ \hat M(\textbf{x})$ represents the (smeared) mass density operator 
\begin{equation}
    \hat M (\textbf{x}) = \frac{1}{(\sqrt{2 \pi} r_C)^3 } \sum_i m_i \sum_s \int d^3 \textbf{z} \: e^{-\frac{(\textbf{z}-\textbf{x})^2}{2 r_C^2} } \hat{\Psi}^{\dagger}_i (\textbf{z},s) \hat{\Psi}_i (\textbf{z},s);
\end{equation}
the operators $\hat{\Psi}^{\dagger}_i (\textbf{z},s)$ and $ \hat{\Psi}_i (\textbf{z},s)$ denote respectively the creation and annihilation operators of a particle with spin $s$ of type $i$ in the point $\textbf{z}$.  

The collapse model, as characterized by Eq. \eqref{CSLmaster}, contains two parameters: The collapse rate $\lambda$, which sets the strength of the collapse, and a noise correlation length $r_C$, which measures the spatial resolution of the collapse.  

The most common theoretical proposal for the numeric value of the parameters is given by Ghirardi, Rimini and Weber \cite{GRW86,GRW87}, these are $\lambda=10^{-16}\text{s}^{-1}$ and $r_C=10^{-7}$m. On the other hand, experiments from gravitational wave detectors and spontaneous X-ray emissions have strongly bounded the parameters of the model,  leaving the region of parameter space around $r_C \sim 10^{-7} - 10^{-4}$m and $\lambda \sim 10^{-18} - 10^{-12}$ s$^{-1}$ viable \cite{BassiNat,Altamura2024}.   

As mentioned in the Introduction, a particular prediction of the CSL model (as well as any collapse model) is spontaneous heating due to the interaction of the system with the noise field; this effect can also be used to constrain the model parameters. Intuitively, one can understand this effect as follows. For a system of particles, the noise field continuously acts on them, causing their wave functions to localize in space spontaneously over time. Given Heisenberg's uncertainty principle, a narrow localization in space implies an increase in kinetic energy, leading to a measurable increase in the system's temperature. In particular, for a body of total mass $M$, modeled as a group of particles, the mass-proportional CSL model with white noise predicts an energy gain given by \cite{Pearle1994}.

\begin{equation}
    \frac{d E}{ dt} = \frac{3 \hbar^2 \lambda M}{4 m_0^2 r_C^2},
     \label{PCSL}
\end{equation}
where $\lambda$ and $r_C$ are the usual parameters of the CSL model, and  $m_0$ is the nucleon mass.

The heating effect has been analyzed for various physical systems, including mono- and multi-atomic crystals \cite{mohammad2018}, phononic modes in matter \cite{vinante2018}, and ultra-cold atoms \cite{Pearle2014}, all resulting in the same prediction as Eq. \eqref{PCSL}. Furthermore, in Refs. \cite{Adler2019, Tilloy2019}, the CSL-induced heating was examined for Fermi liquids;  the energy gained in such a system also coincides with Eq. \eqref{PCSL}. The calculations derived in \cite{Adler2019, Tilloy2019} were applied to the case of neutron stars.

It is important to mention that, although  Eq. \eqref{PCSL} was originally derived considering a Fermi liquid, the calculation is valid for any system of fermionic particles. Moreover, the derivation does not employ any particular statistical distribution. This implies that Eq. \eqref{PCSL} can be applied to a wide variety of astrophysical objects, such as white dwarfs, planets, or other compact objects, rather than being restricted solely to neutron stars. This approach will be adopted in the subsequent sections of our paper.

\section{Bounds from white dwarfs astrophysics} \label{sec3}
White dwarfs are the final stage in the evolution for the majority of stars in the Universe \cite{Isern1998,2010A&ARv..18..471A}. They are compact dense objects with masses similar to the Sun and radii similar to the Earth. Their structure is sustained by the pressure of the degenerate electrons and, since they cannot obtain energy from nuclear reactions, their evolution can be described primarily as a cooling process \cite{Mestel1952}. Their relatively well known physics and the possibility to actually measure their cooling speeds makes them ideal for testing a huge variety of physical models \cite{Isern2008ApJ,Corsico2012,M3B2014JCAP,M3B2014A&A,Bottaro2023}. 
There are two main, mostly independent, ways to measure the
cooling rate of WDs.  One is based on the secular drift of the
pulsation period of individual white dwarf variables, and the
other is based on the number counts of white dwarfs at different
luminosities in a given stellar population (the so-called white
dwarf luminosity function). In this Section we will make use of these two tools in addition to the stellar parameters measured for individual WDs to obtain three different bounds for the CSL parameters.

\subsection{Bounds from the thermal balance of individual white dwarfs} \label{secthermal}
Following  \cite{Adler2019}, we can estimate the value of $\lambda/{r_C}^2$ at which the anomalous CSL heating would start to produce observable consequences in WDs by balancing the energy radiated away from the star with that produced by the anomalous CSL heating effect ($L_{\rm CSL}$). Assuming that neutrino emission is negligible, and that white dwarf losses its energy by thermal radiation emission from the photosphere (i.e. the stellar luminosity $L_\star$), we have
\begin{equation} \label{eq31}
L_\star=4\pi\sigma {R_\star}^2 T_{\rm eff}^4=L_{\rm CSL}=\frac{3 \hbar^2}{4 {m_0}^2}\frac{\lambda}{{r_C}^2} M_\star,
\end{equation}
where $R_\star$ is the photospheric radius, $T_{\rm eff}$ the effective temperature, and $M_\star$ the mass of the star, $\hbar$ the reduced Planck's constant, $\sigma$ the Stefan-Boltzmann's constant, and $m_0$ is the atomic mass unit.

From Eq. \ref{eq31}, we find that values of $\lambda/{r_C}^2$ such that
\begin{equation}
\frac{\lambda}{{r_C}^2} > \left. \frac{\lambda}{{r_C}^2}\right|_{\rm eq} = \frac{4 L\star m_0^2}{3\hbar^2 M_\star}
\label{eqparam}
\end{equation}
would lead to a net heating of the star, as the energy radiated away from the photosphere would not be enough to counteract the anomalous CSL heating.

Now, we need a sample of stars to test this expression. Since we are interested in obtaining the lowest bound possible, we will focus on stars that are as massive and faint as possible. A good sample can be found in \cite{Bergeron2022}, where we can use the masses and luminosities in Table 4. By applying Eq. (\ref{eqparam}) to those stars, we can obtain the equilibrium values for the CSL parameters for the whole sample. The lowest numerical bounds are obtained for J1251+4403 and J1220+0914, with $1\sigma$ and $3\sigma$ precision values\footnote{ $1\sigma$- and $3\sigma$-values were derived by allowing $1\sigma$ and $3\sigma$ changes in the reported stellar parameters in the least favorable direction.} of:
\begin{eqnarray}
    (\lambda/r_C^2)_{\text{J1251+4403},1\sigma} & = & 4.700\times 10^5\text{s}^{-1}\text{m}^{-2}, \nonumber \\
    (\lambda/r_C^2)_{\text{J1251+4403},3\sigma} & = & 4.766\times 10^5\text{s}^{-1}\text{m}^{-2} . \nonumber \\
    (\lambda/r_C^2)_{\text{J1220+0914},1\sigma}  & = & 5.931 \times 10^5\text{s}^{-1}\text{m}^{-2} \nonumber \\
    (\lambda/r_C^2)_{\text{J1220+0914},3\sigma}  & = & 6.019 \times 10^5\text{s}^{-1}\text{m}^{-2} \nonumber
    \end{eqnarray}

This constraint is more than two orders of magnitude more restrictive than the one obtained in \cite{Adler2019}, which was $\lambda/r_C^2 = 9.43\times 10^7\text{s}^{-1}\text{m}^{-2}$ corresponding to the neutron star PSR J1840-1419. 


The bounds of $\lambda/{r_C}^2$ derived from the thermal balance of WDs are, however, less restrictive than the values derived from icy giants in the Solar System \cite{Adler2019}. 
It should be noted, however, that the use of Eq. \ref{eq31} for the planets for the Solar System is questionable at best since, for all of them, the radiation coming from the Sun cannot be neglected in the energy balance of the planet.
In Section \ref{sec_planet_revisited}, the bounds that can derived from the anomalous CSL heating of the planets in the Solar System will be explored.

However, bounds relying on the parameters of individual studies of stars should be taken with caution, mainly because they depend on the accuracy of the stellar parameters determined for each star. Interestingly, as mentioned in the introduction, there are two main independent methods to measure the cooling of white dwarfs. These observable properties can then be used to derive accurate bounds on the value of $\lambda/r_C^2$.


\subsection{Bounds from the secular period drift of variable white dwarfs}

The first approach we will discuss comes from the period drift of variable white dwarfs. In what follows, we will provide a simple estimation of the effectiveness of this approach in the context of anomalous CSL heating.

It is possible to show from relatively simple arguments \citep{1983Natur.303..781W} that temporal changes in the periods of normal modes in white dwarfs ($\Pi$, $\dot{\Pi}$) are related to the WD cooling speed of the core ($\dot{T}_c$), and the speed at which the envelope contracts ($\dot{R_\star}$), as
\begin{equation}
\frac{\dot{\Pi}}{\Pi}=-a\,\frac{\dot{T}_c}{T_c}+b\,\frac{\dot{R_\star}}{R_\star},
\end{equation}
where $a$ and $b$ are two positive constants that depend on the specific mode and evolutionary state of the white dwarf. For cold low-luminosity white dwarfs, the contraction term is almost negligible \citep{1992ApJ...392L..23I} $b\simeq 0$, and the period drift of normal modes is dominated by cooling. For stars that are sufficiently cold (old) that neutrino emission is negligible but still hot (young) enough that crystallization has not yet occurred, the surface luminosity of the star can be equated to the loss of thermal energy from the ions, given by $L_\star \simeq - \langle c_v\rangle M_\star \dot{T}_c$ \citep{2010A&ARv..18..471A}.

This is the regime in which the most numerous white dwarf variables are found ---the so-called DAVs or ZZ Ceti stars, see \cite{2019A&ARv..27....7C}. Any additional cooling or heating mechanism will increase of decrease the period drift. Perturbatively, we can derive that an anomalous heating ($L_{\rm CSL}$) will lead to a change in the rate of period drift of DAV white dwarfs ($\delta \dot{\Pi}$) given by (see \cite{1992ApJ...392L..23I}), $L_{\rm CSL}/L_\star\simeq \delta{\dot{\Pi}}/\dot{\Pi}$.

Currently, there are only three DAV stars for which the period drift has been measured (G117-B15A, R548, and  L19-2 \cite{2019A&ARv..27....7C}). In particular, the period $\Pi\simeq 215$ s  of G117-B15A has been found to be extremely stable. More than 40 years of measurements of this period has allowed an unprecedented precision ($\Pi=215.19738823\pm0.00000063$ s), allowing the most precise determination of the period drift in any DAV ($\dot{\Pi}=(5.47\pm 0.82) \times 10^{-15}$ s s$^{-1}$), see \cite{2021ApJ...906....7K}. This corresponds to a precision of $\delta\dot{\Pi}/\dot{\Pi}\simeq 0.15$. This implies that only an anomalous heating of about $L_{\rm CSL}/L_\star\simeq \delta{\dot{\Pi}}/\dot{\Pi}\simeq 0.15$ would be detectable. For the luminosity and mass of G117-B15A ($\log{L_\star/L_\odot\simeq -2.5}$ and $M_\star\simeq 0.59 M_\odot$ \cite{Corsico2012}) this implies that, currently, the best possible constraints to the CSL anomalous heating coming from variable white dwarfs would be close to $\lambda/{r_C}^2 \lesssim 5.2\times 10^7$ m$^{-2}$s$^{-1}$. This value is already smaller than the estimation coming from the thermal balance in the neutron star PSR~J1840-1419 found in Ref. \citep{Adler2019}. As a matter of fact, values of $\lambda/{r_C}^2 \gtrsim 5.2\times 10^7$ m$^{-2}$s$^{-1}$ would contradict the actual cooling speed measured in  G117-B15A \cite{2021ApJ...906....7K}. While this is the best bound that can be obtained from the asteroseismology of DAV WDs, similar bounds could be obtained from the other two DAV WDs for which the period drift was actually measured  (R548, and  L19-2 \cite{2019A&ARv..27....7C}).

\subsection{Bounds from the White Dwarf Luminosity Function (WDLF)}

The next approach we will take involves exploring the WDLF, defined as  the number of white dwarfs per cubic parsec and unit bolometric magnitude as function of the bolometric magnitude \cite{2010A&ARv..18..471A,M3B2014JCAP,GarciaBerro2016}. If we limit ourselves to the intermediate luminosity regime, where both crystallization and neutrino emission can be neglected, and we assume a typical mass of $M_{\rm WD}$ for the typical WDs, the WDLF can be shown to be
\cite{Bottaro2023}:

\begin{equation} \label{eq34}
    \frac{dN}{dM_{\text{bol}}} = B_3\ 2.2 \times 10^{-4} \frac{10^{-4M_{\text{bol}}/35}L_{\odot}}{78.7 L_{\odot}10^{-2M_{\text{bol}}/5}+L_X} \bigg( \frac{M_{\rm WD}}{M_\odot}\bigg)^{5/7} \bigg( \sum_j \frac{X_j}{A_j}\bigg) \text{pc}^{-3}\text{mag}^{-1},
\end{equation}
where $L_X$ represents the additional contribution, which can be either extra cooling by emission of particles \cite{Bottaro2023} or, in our work, the CSL heating mechanism. $B_3$ is the constant birthrate normalised to $10^{-3}\text{pc}^{-3}\text{Gyr}^{-1}$, $X_j$  is the mass fraction of the element $j$ and $A_j$ the mass (nucleon) number of the element $j$. In the present work we will consider an equal mixture of carbon and oxygen and a typical WD mass of $M_{\rm WD}=0.6 M_\odot$.

Introducing the CSL heating mechanism from Eq. \ref{PCSL} in Eq. \ref{eq34} we obtain the following expression:

\begin{equation}
    f_{\text{th}}\equiv \bigg(\frac{dN}{dM_{\text{bol}}}\bigg)_{\lambda/r_C^2} =\frac{B_3\ 2.2 \times 10^{-4} 10^{-4M_{\text{bol}}/35}L_{\odot}}{78.7 L_{\odot}10^{-2M_{\text{bol}}/5}-\frac{3\hbar\lambda M_{\rm WD}}{4 m_0^2r_C^2}} \bigg( \frac{M_{\rm WD}}{M_\odot}\bigg)^{5/7} \bigg( \sum_j \frac{X_j}{A_j}\bigg) \text{pc}^{-3}\text{mag}^{-1},
    \label{eq:WDLF}
\end{equation}
where the minus sign in the CSL term in the denominator comes from the fact that this mechanism is heating the star. With the previous assumptions, if we consider $\lambda/r_C^2 = 0$ we recover Mestel's cooling law \cite{Mestel1952} (scaled with $B_3$). The extra CSL term introduces a divergence in the WDLF, predicting a limit magnitude at which the WDs should accumulate and could not cool anymore. This divergence occurs at higher luminosities for higher values of $\lambda/r_C^2$. Note that for sufficiently high values for the CSL parameters, there are actually fainter stars in the observable WDLF and, since the model predicts a brighter limit magnitude, these values are already discarded.

For the following test, we will estimate the values for $\lambda/r_C^2$ for which the separation between their predicted WLDF and the observable data is sufficiently large that we can discard them using a $\chi^2$ statistic. We will concentrate in the intermediate luminosity range $10.5<M_{\text{bol}}< 15$ where both the energy release by crystallization and neutrino cooling can be neglected, allowing the use of Eq. \ref{eq:WDLF}. To define the value of $B_3$, we will ensure that the theoretical WDLF yields the same number of stars per pc$^3$ as the observational WDLF, by imposing

\begin{figure}[t!] 
    \centering
    \includegraphics[scale=0.6]{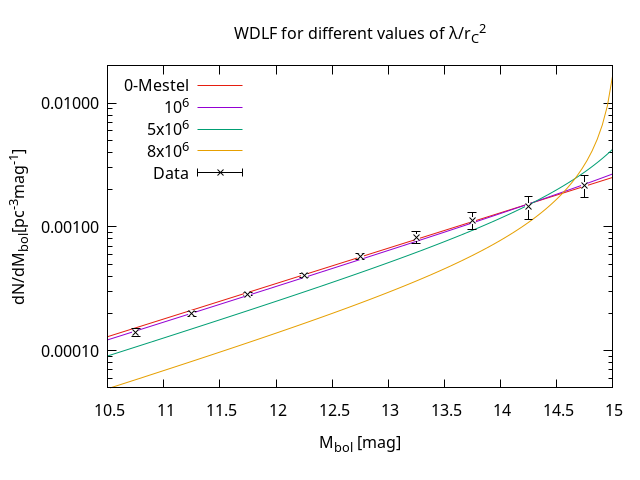}
    \caption{Different models compared with the data of \cite{M3B2014A&A}. At higher values for $\lambda/r_C^2$ the thoretical WDLF diverges at lower magnitudes. All functions were normalized to have the same area under their curves on the interval displayed.}
    \label{figWDLF}
\end{figure}
\begin{equation}
    \int_{10.5}^{15}f_{\text{th}} dM_{\text{bol}} = \sum_{10.5}^{15} f_{\text{obs}} \Delta M_{\text{bol}} ,
\end{equation}
where $f_{\text{obs}}$ is the observational data given in \cite{M3B2014A&A} and shown in Figure \ref{figWDLF} and $\Delta M_{\text{bol}}=0.5$ is the width of every box in the figure. This normalization means that all WDLF will count the same number of stars in the given interval and this will match the observed data. This is a constrain that will remove a degree of freedom in the following analysis.

\begin{figure}[t!]
    \centering
    \includegraphics[scale=0.6]{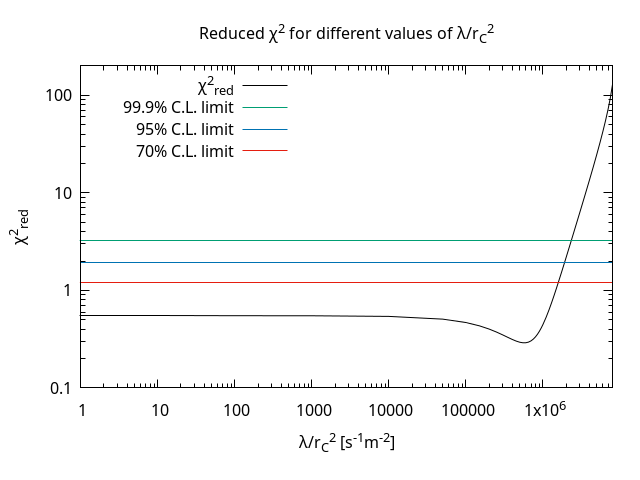}
    \caption{The reduced $\chi^2$  statistical for the CSL parameters and their intersection with the limits of different confidence levels. All CSL parameters higher than these intersections are discarded with the corresponding confidence level. }
    \label{figchi}
\end{figure}

Assuming that the WDLFs and the observed data are realizations of distributions with the same mean value $\mu_i$ at each $M_{\text{bol,}i}$ and Gaussian errors, as in \cite{M3B2014A&A}, then the quantity $\chi^2_{\text{red}}$ 
\begin{equation}
    \chi^2_{\text{red}} (\lambda/r_C^2) = \frac{1}{8} \sum_{M_{\text{bol,}i}=10.5}^{15} \frac{f_{\text{obs}}( M_{\text{bol,}i})-f_{\text{th}}( M_{\text{bol,}i})}{\sigma(M_{\text{bol,}i})^2},
\end{equation}
will follow a $\chi^2$ distribution and we can use it to discard the values of the CSL parameters that yield a $\chi^2_{\text{red}}$ higher than the limit given by certain confidence level. In this expression the factor $1/8$ comes from the fact that we need to divide the $\chi^2$ to the number of deegres of freedom of the sample to obtain the reduced $\chi^2_{\text{red}}$. 
We plot the resultant $\chi^2_{\text{red}}$ in Figure \ref{figchi}.

Depending on the confidence level pursued, we can obtain different bounds for the CSL parameters, as shown in Figure \ref{figchi}. These bounds are, nevertheless, all of the same order of magnitude:
\begin{eqnarray}
    \frac{\lambda}{r_C^2}\bigg |_{70\% \ \text{C.L.}} &\approx& 1.576 \times 10^6\text{s}^{-1}\text{m}^{-2} ,\nonumber
    \\
    \frac{\lambda}{r_C^2}\bigg |_{95\% \ \text{C.L.}} &\approx& 1.912 \times 10^6\text{s}^{-1}\text{m}^{-2} , \nonumber
    \\
    \frac{\lambda}{r_C^2}\bigg |_{99.9\% \ \text{C.L.}} &\approx& 2.340 \times 10^6\text{s}^{-1}\text{m}^{-2} . \nonumber
\end{eqnarray}


In Section \ref{discussion} we will discuss how these bounds could be further improved and compare them with the obtained in Section \ref{sec_planet_revisited}.

\section{Bounds from the Solar System}
\label{sec_planet_revisited}
\begin{table}[] 
\centering
\caption{}
\begin{tabular}{llll}
\hline
Planet    & Intrinsic heat flux{[}W/m$^2${]} & $(\lambda/r_C^2)_{1\sigma}[\text{s}^{-1}\text{m}^{-2}]$ & $(\lambda/r_C^2)_{3\sigma}[\text{s}^{-1}\text{m}^{-2}]$ \\ \hline
Jupiter & $7.485\pm0.160$                  & $8.676\times 10^4$                                      & $9.040\times 10^4$                                        \\
Saturn  & $2.010\pm0.140$                  & $5.447\times 10^4$                                      & $6.155\times 10^4$                                        \\
Uranus  & $0.042^{+0.047}_{-0.042}$        & $2.781\times 10^3$                                      & $5.717\times 10^3$                                        \\
Neptune & $0.433\pm0.046$                  & $1.195\times 10^4$                                      & $1.425\times 10^4$                                        \\ \hline
\end{tabular}
\label{tablegas}
\end{table}

As we commented in Section \ref{sec3}, the radiation coming from the Sun cannot be neglected in the energy balance of the planets in the Solar System. This is particularly true for planets such as Earth or Uranus, for which the radiation absorbed from the Sun is very close to that emitted from the surface. Moreover, each planet has an albedo that depends on the characteristics of its surface and atmosphere, and therefore only absorbs a fraction of radiation. Moreover, other effects such as the presence of an atmosphere (and therefore greenhouse effects) can increase the surface temperature beyond the value of the effective temperature. By using an equation like Eq. \ref{eq31} together with the black-body temperatures obtained from \cite{Williams2016}, the estimation performed by \cite{Adler2019} effectively compares the anomalous CSL-heating with the fraction of the Solar luminosity that is absorbed (i.e. not reflected) by each planet and needs to be thermally re-emitted in equilibrium\footnote{This is because the black-body temperatures presented in \cite{Williams2016} only account for the thermal re-emission of the energy absorbed from the Sun, neglecting intrinsic planetary heat sources \citep{2010ppc..book.....P}.}. Note that this is not correct, the hypothetical CSL heating should either be compared to the total observed luminosity of the planet or directly compared to the intrinsic heat released by the planet once the known solar contribution is removed. Moreover,
 using Eq. \ref{eq31} squanders a great opportunity to put stricter bounds, as in many cases most of the energy radiated away from a planet comes directly from the Sun and can be easily removed from the energy budget. It is the intrinsic power of the planet, i.e. the difference between the total incoming energy from the Sun and the total emitted radiation, what should be used to set constraints on the anomalous CSL-heating ---see Ref. \cite{2011A&A...529A.125K} for a similar analysis in the case of the hypothetical heating introduced by a cosmological variation of the fine structure constant.

 Considering again that CSL heating $P_\text{CSL}$ equals the power heat emitted from their interior $P_\text{int}$, we can obtain the corresponding bounds to the parameters:

 \begin{equation} \label{eqplanet}
     \frac{\lambda}{r_C^2} = \frac{4m_0^2P_\text{int}}{3\hbar^2 M_\text{p}} ,
 \end{equation}
with $M_\text{p}$ being the mass of each planet.
\subsection{Giant planets}

First, we will consider the case of the gas and icy giants, for which the energy balance was determined by Voyager and Cassini \cite{1968ApJ...152..745H, 1991JGRS...9618921P, 1990Icar...84...12P, 2018Nature}. Their main source of internal energy is the gravothermal contraction of the planets \cite{2015trge.book..529G}. By comparing the observationally derived internal heat flow to the potential anomalous CSL heating power, we can derive the values of $\lambda/r_C^2$ beyond which the CSL heating power would be sufficient to produce the entirety of the observed heat flux. This assumption, i.e. that the observed heat flow represents the mean steady-state value of the internal heat flow, sets a strict constraint on the values of $\lambda/r_C^2$.  Applying Eq. \ref{eqplanet} to the internal heat flow plus its $1\sigma-$ and $3\sigma-$errors, we obtain the corresponding bounds to the parameters from each planet (Table \ref{tablegas}). As expected the derived values are smaller than those derived by \cite{Adler2019}, the only exception being Neptune for which the  power absorbed from the Sun 
is smaller than the actual measured internal heat flux (see table 2 of \cite{2015trge.book..529G}). Consequently,  the use of the black-body temperature of \cite{Williams2016} in Ref. \cite{Adler2019} underestimates the heat emitted from the planet leading to an overly optimistic bound.



\subsection{Earth and Moon}

Now we turn our attention to our planet and the Moon. Naturally, the heat flow on Earth has been  measured more precisely than on the giant planets.  One of the latest determinations was performed in Ref. \cite{2010EGUGA..12.3303D}. The authors provide a revised global estimate of the Earth surface heat flow which is based in a huge  data set of 38347
measurements distributed over the whole surface of the Earth (both in land and sea). The global estimated value of the surface heat flow in Ref. \cite{2010EGUGA..12.3303D} is $47\pm 2$ TW. From Eq. \ref{eqplanet} this leads to the following bounds for the CSL parameters:
\begin{eqnarray}
    (\lambda/r_C^2)_{\oplus,1\sigma} &=& 2.751\times 10^3 \text{s}^{-1}\text{m}^{-2} , \nonumber \\
     (\lambda/r_C^2)_{\oplus,3\sigma} &=& 2.975 \times 10^3 \text{s}^{-1}\text{m}^{-2} . \nonumber 
\end{eqnarray}
This bound has the advantage of relying on experiments conducted on our own planet and covering the entire surface of the Earth. The techniques involved are also more precise and sophisticated than those possible for other celestial bodies.

On the other hand, the Moon had two \textit{in situ} experiments, Apollo 15 and Apollo 17, where the heat flow was estimated between 12 and 18 mW/m$^2$ \cite{LHFE}. However, newer studies suggest that these values are overestimated by the influence of other mechanisms \cite{2023FrASS..1079558W} and estimate a significantly lower value of $4.9\pm0.2$mW/m$^2$, using Chang’E-2 microwave radiometer data and Diviner observations. With these new estimations, we obtain the lowest bound for the CSL parameters in this work:
\begin{eqnarray}
    (\lambda/r_C^2)_{\leftmoon,1\sigma} &=& 8.840\times 10^2 \text{s}^{-1}\text{m}^{-2} , \nonumber \\
     (\lambda/r_C^2)_{\leftmoon,3\sigma} &=& 9.533\times 10^2 \text{s}^{-1}\text{m}^{-2} . \nonumber
\end{eqnarray}


In Section \ref{discussion} we will compare these bounds with the previously found in Section \ref{sec3} and by other experiments. Also, we will discuss how they could be improved.

\section{Estimation using PSR J2144–3933} \label{secneutron}

We can revisit the astrophysical bound given by neutron stars to the CSL parameters by following the same approach done in Section \ref{secthermal}, that is, finding the coldest neutron star in order to take advantage of the strong dependence of $L_\star$ with the effective temperature. Again considering that the object is evolving through a simple cooling process and the star losses its energy by Stefan-Boltzmann's law we obtain, as in \cite{Adler2019}:

\begin{equation} 
    \frac{\lambda}{r_C^2} = \frac{16\pi \sigma m_0^2 R_\star T_{\text{eff}}^4}{3\hbar^2M_\star} .
\end{equation}

In \cite{2019ApJ...874..175G} the authors estimate an upper bound for the surface temperature of PSR J2144–3933, obtaining that it can not be higher than $T_{\text{eff}}<42000$K. This would be, then, the coldest neutron star observed to this day. Assuming the least favorable values for the mass $M_\star=1M_\odot$ (lowest) and a coordinate radius of $R_\star=16$km (highest possible) \citep{stellarcollapse, Freire2024}, we obtain a value for the CSL parameters of $\lambda/r_C^2=9.51\times10^4\text{s}^{-1}\text{m}^{-2}$. This value is three orders of magnitude lower than the one obtained in \cite{Adler2019} for PSR J 1840-1419. 


Similarly to the case discussed in Section \ref{secthermal}, higher values for the CSL parameters would result in a net heating of the star. However, since this bound is again obtained from data of an individual object, we face the same limitation of dependence on the accuracy in the determination of the stellar parameters. This situation is exacerbated by the fact that the stellar parameters of PSR J2144-3933 are not even determined; it might even be possible that the object is not a neutron star. For these reasons, and although this value serves as an estimation, we will not consider it a reliable bound like those previously determined for other astrophysical objects.

\section{Discussion and conclusions} \label{discussion}

\begin{table}[] 
\centering
\caption{Bounds with $3\sigma$ precision obtained for the CSL parameters in the different astrophysical approaches.}
\begin{tabular}{ll}
\hline
Method used                     & $\lambda/r_C^2[\text{s}^{-1}\text{m}^{-2}]$ \\ \hline

White dwarf luminosity function & $2.340 \times 10^6$                           \\
Individual white dwarf data     & $4.766 \times 10^5$                           \\

Icy giants heat flux                     & $5.717\times 10^3$                                         \\
Earth heat flux                      & $2.975\times 10^3$
\\
Moon heat flux & $9.533\times 10^2$ \\
\hline
\end{tabular}
\label{tablefinal}
\end{table}

\begin{figure}
    \centering
    \includegraphics[scale=0.6]{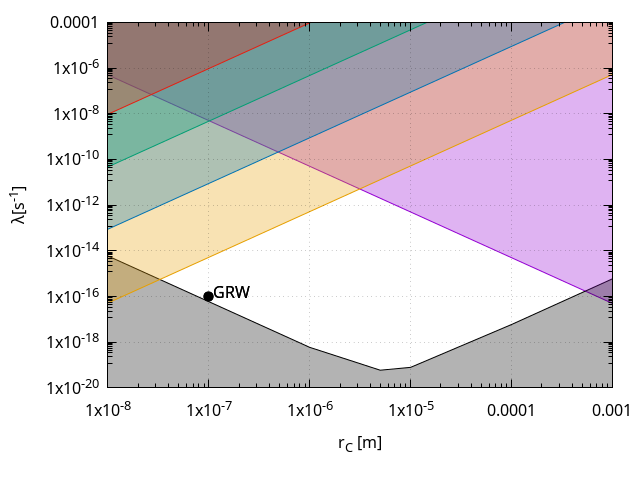}
    \caption{Schematic exclusion plot for the CSL parameters. The grey area is a theoretical limit excluded from the requirement that macroscopic superpositions do not persist in time \cite{Toros2017} (plot obtained from \cite{Altamura2024}). The red zone is the bound found by Adler in \cite{Adler2019} for the CSL anomalous heating of a neutron star. The green zone is our bound found by thermal emission of the white dwarf J1251+4403 and the blue zone is from the heat flow in the Moon. For the sake of completeness, we have added two more exclusion zones. The purple is from LISA Pathfinder \cite{PhysRevD.95.084054}, and the yellow zone is from Majorana Collaboration  \cite{Majorana2022} (plots obtained from \cite{Altamura2024}).  The white region is yet to be determined. Also the theoretical GRW values are shown.}
    \label{figcotas}
\end{figure}

In this work we explored the consequences of the anomalous CSL heating in a variety of astrophysical objects. This novel predicted effect enables us to constrain the characteristic parameters of the model by estimating the values at which CSL heating would be significant enough to contradict the observational data. The results with $3\sigma$ precision are listed in Table \ref{tablefinal}.

Based on the preceding analysis, we can now briefly discuss the advantages and disadvantages of the observational bounds found in each case.

White dwarfs offer probably the most robust bounds due to their relative simplicity, well known physics, and huge amount of observational data available. However, the most important feature of these objects is that we can actually measure their cooling using the white dwarf luminosity function (WDLF) and the period drift of variable white dwarfs. Most importantly the WDLF allows us to measure the time-averaged cooling rate, over long timescales, which is precisely the aspect that would be influenced by the  anomalous CSL heating.
As the WDLF also relies on the measurements of thousands of WDs, its constraints are not affected by the uncertainties in determining the stellar parameters of individual stars.  Moreover, in this work, we adopted a very simple model for the WDLF, which limited us to utilizing the intermediate luminosity region of the WDLF ($10.5<M_{\text{bol}}< 15$). Current constraints could be enhanced by incorporating the new GAIA-derived WDLFs \cite{2021MNRAS.502.1753T}, as well as employing more sophisticated stellar and population models to derive the theoretical WDLF \cite{2023A&A...677A.159T}.

Aside from the WDLF, the detection of ultra-cool white dwarfs \cite{Bergeron2022} provides a stringent constraint on the CSL parameters due to spontaneous heating. Given their short reaction timescales (thermal, hydrostatic), long-term fluctuations in the heat flow are not expected and should not affect the constraints obtained from eq. \ref{eqparam}. The main drawback of estimations based on the surface parameters of individual WD is their strong dependence on the accuracy of these determinations. As an example, let us note that for the star producing our best bound (J1251+4403), previous studies \cite{2015MNRAS.449.3966G} indicated a much lower effective temperature and mass. Interestingly, current models suggest that stars used in this work to better constrain CSL-heating are only a few Gyrs old. In the future, much older, and consequently dimmer WDs will be detected. Constraints derived from ultra-cool white dwarf will be improved as dimmer and dimmer WDs are detected in the solar neighbourhood \cite{2024A&A...683A..33G}.

Despite not considering the estimated value in Sect. \ref{secneutron} a reliable bound to the CSL parameters, it is worth mentioning its potential for future considerations.  If a neutron star is detected with an effective temperature comparable to that of a typical white dwarf, the constraint that could be obtained would be significantly lower due to the much more compact size of neutron stars.

The Solar System yields lower numerical values for the bounds, primarily due to the low intrinsic power emitted by the planets. Specifically, Uranus provides the lowest bound among the giant planets, and this could potentially decrease further with more precise estimations of its heat flow. Notably, this bound could even be lower than the one obtained from Earth. The main shortcoming is that these values rely on the implicit assumption that the measured heat flux is representative of its mean value over time.
For example, fluctuations in the heat flow have been detected on Saturn \cite{2015GeoRL..42.2144L}, which suggests the actual bound could be higher if the power emitted increases over time. Another shortcoming is the lack of more recent observations, especially for the icy giants.

Earth warrants special consideration. Given the exceptionally high precision of heat flow measurements, which are based on tens of thousands of individual measurements covering the entire surface, Earth allows for a much stronger and more robust constraint compared to the giant planets. 

Finally, the Moon yields the lowest bound in this work, which is five orders of magnitude lower than that previously found in \cite{Adler2019} for neutron stars, more than two orders of magnitude below our lowest bound for white dwarf astrophysics, and lower than the value obtained for Earth by a factor of 3. However, as promising as this bound is, one should keep in mind that it relies on measurements conducted in a limited region of the Moon. Consequently, possible spatial and temporal fluctuations of the heat flux could lead to variations in the observational estimation of the mean heat flow. Similar to the bounds obtained from the gas and icy giants, this calculation assumes a stationary state for their intrinsic power, which cannot be assured due to the impossibility of performing experiments over the long characteristic timescales of these types of celestial bodies.

As mentioned in Sect. \ref{sec:intro} and Sect. \ref{sec2}, the CSL parameters have been constrained by other experiments. In Figure \ref{figcotas}, we plot the two most restrictive ones obtained from gravitational waves by LISA Pathfinder \cite{PhysRevD.95.084054}  and from spontaneous X-ray emission by the Majorana Collaboration \cite{Majorana2022}, alongside our astrophysical bounds derived from this work.   In particular, the Majorana Collaboration obtained a value for $(\lambda/r_C^2)_\text{Majorana} = 4.94\pm 0.15 \times 10^{-1}\text{s}^{-1}\text{m}^{-2}$, which is more than three orders of magnitude lower than our lowest bound.  The spontaneous X-ray emission is predicted by the CSL model from the interaction between the noise field and charged particles \cite{Fu1997}. The proposed theoretical value of $(\lambda/r_C^2)_\text{GRW}\simeq 0.01 \: \text{s}^{-1}\text{m}^{-2}$ remains within the acceptable region of parameter space.

On the other hand, the astrophysical bounds derived in this work, based on the spontaneous heating effect, are the strongest of this kind to date. In particular, the parameter constraints from the intrinsic heat flux of the Earth and Moon establish $\lambda \leq 2.975 \times 10^{-11}$ s$^{-1}$ and $\lambda \leq 9.533 \times 10^{-12}$ s$^{-1}$, respectively, for the reference value of $r_C = 10^{-7}$ m. These values of the CSL model parameters (characterized by white noise type) are currently the lowest obtained using astronomical and cosmological observations (see Table 1 of \cite{BassiNat}). Furthermore, the Earth/Moon-derived bounds are approximately three orders of magnitude lower than those from cold atom experiments \cite{Kovachy2015} and comparable to the order of magnitude established from low-temperature experiments of phonons \cite{vinante2018, mohammad2018}.

Our astrophysical bounds from spontaneous CSL heating could be further improved by testing on other planets or minor bodies in the Solar System. In particular, terrestrial planets with little geological activity such as Mars or Venus, or even dwarf planets and asteroids, could provide lower values due to their lack of an internal heat source. The shortcoming of this approach is the difficulty of obtaining reliable or even \textit{in situ} measurements of the heat flow for these celestial bodies.








\acknowledgments
MMO and M3B are partially supported by PIP-2971 from CONICET (Argentina) and by PICT 2020-03316 from Agencia I+D+i (Argentina). G.L. is supported by grants Universidad Nacional de La Plata I+D G175, and PIP11220200100729CO CONICET, and CONICET (Argentina). 

\bibliographystyle{JHEP}
\bibliography{cslheating.bbl}




\end{document}